\documentclass{PoS}
\usepackage{units}
\usepackage{comment}

\title{Search for Ultra High Energy Cosmic Rays from Space - The JEM-EUSO 
Program}

\ShortTitle{The JEM-EUSO Program}

\author{
\speaker{M.~Bertaina}$^1$, 
for the JEM-EUSO Collaboration\footnote{for
collaboration list see PoS(ICRC2019)1177} \\
\llap{$^1$} Department of Physics, University of Torino, Italy\\
E-mail: \email{bertaina@to.infn.it}}

\abstract{
The origin and nature of Ultra-High Energy Cosmic Rays (UHECRs) remain unsolved
 in contemporary astroparticle physics. To give an answer to these questions is
 rather challenging because of the extremely low flux of a few per km$^2$ per 
century at extreme energies (i.e. E $>$ 5$\times$10$^{19}$ eV). The central
objective of the JEM-EUSO program, 
Joint Experiment Missions for Extreme Universe Space Observatory,
is the realisation of an ambitious space-based mission devoted to UHECR 
science. A super-wide-field telescope will look down from space onto the night 
sky to detect UV photons emitted from air showers generated by UHECRs in the 
atmosphere. 
The JEM-EUSO program includes several missions from ground
(EUSO-TA), from stratospheric balloons
(EUSO-Balloon, EUSO-SPB1, EUSO-SPB2), 
and from space
(TUS, Mini-EUSO) employing fluorescence
detectors to demonstrate the UHECR observation from space and
prepare the large size missions K-EUSO and POEMMA.
We review the scientifical objectives associated with the developing projects
of the JEM-EUSO program and the technological achievements allowing them.
}

\FullConference{36th International Cosmic Ray Conference -ICRC2019-\\
		July 24th - August 1st, 2019\\
		Madison, WI, U.S.A.}

\begin{document}

\section{Introduction}
\begin{figure}[t]
\begin{center}
\includegraphics[width=0.8\textwidth]{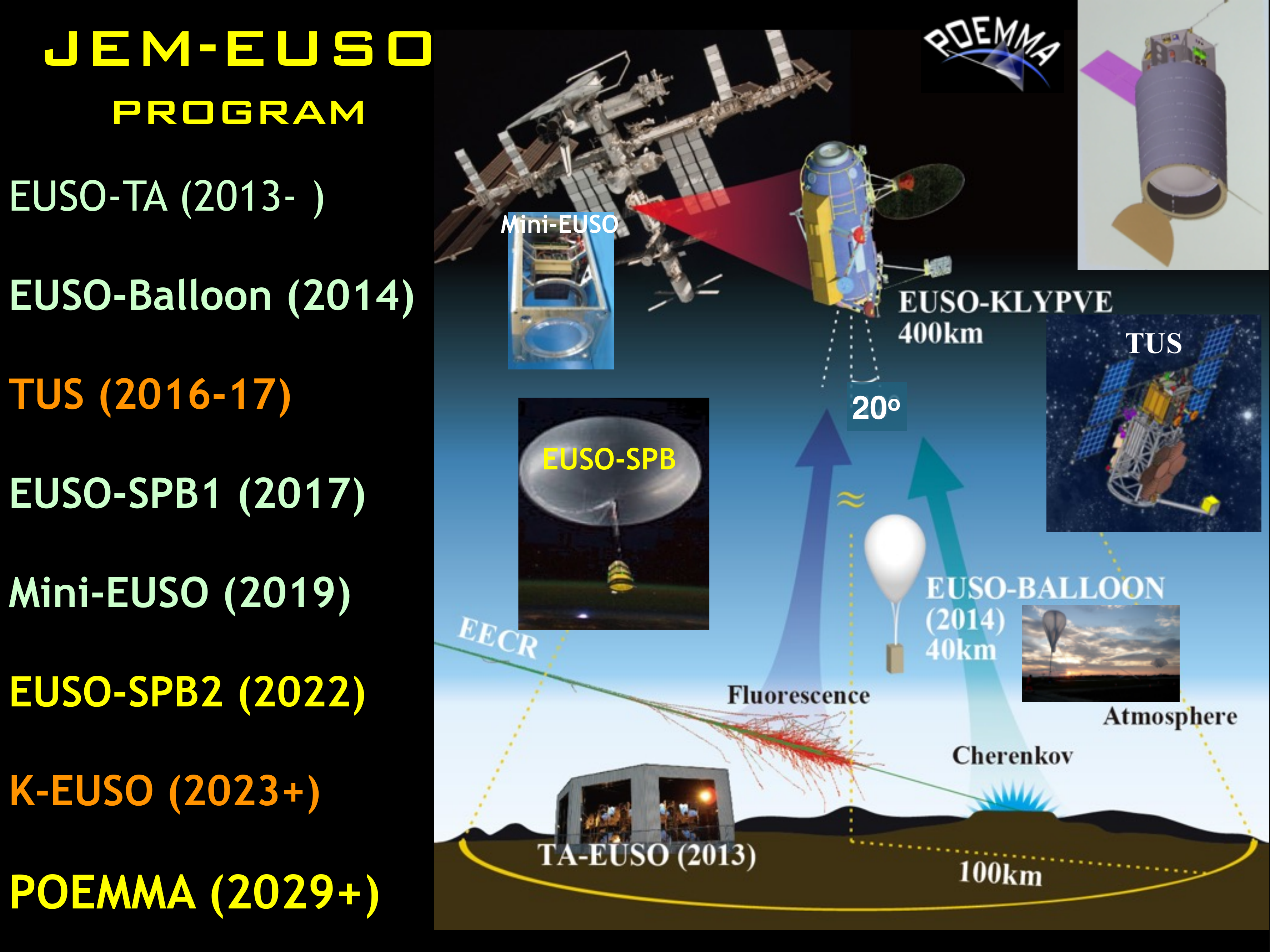}
\end{center}
\caption{Roadmap of the JEM-EUSO program.
See text for details.}
\label{fig:roadmap}
\end{figure}

The main objective of the JEM-EUSO program~\cite{EUSO-Program}, 
Joint Experiment Missions for Extreme Universe Space Observatory, is
the realization of an ambitious space-based mission devoted to scientific 
research of UHECRs~\cite{jemeuso}. 
The JEM-EUSO program includes several missions from ground
(EUSO-TA~\cite{eusota}), from stratospheric balloons
(EUSO-Balloon~\cite{eusobal}, EUSO-SPB1~\cite{spb1}, EUSO-SPB2~\cite{spb2}), 
and from space
(TUS~\cite{tus}, Mini-EUSO~\cite{minieuso}) employing fluorescence
detectors to demonstrate the UHECR observation from space and
prepare the large size missions K-EUSO~\cite{keuso} and POEMMA~\cite{poemma}. 
The timeline of the different projects is shown in
Fig.~\ref{fig:roadmap}, while the key results obtained so far are summarized in
\cite{fenu_icrc19}.

\section{The ground-based telescope EUSO-TA}
\label{eusota}
EUSO-TA is a ground-based telescope, installed at the TA site in Black Rock
Mesa, Utah, USA. This is the first detector to successfully use a Fresnel lens
based optical system and Multi-Anode Photomultipliers (MAPMT). 
Each MAPMT has 64 channels, and they are grouped in blocks of 2$\times$2 to 
form one Elementary Cell (EC). Nine ECs form one Photo-Detector Module (PDM)
which has, therefore, 2304 channels encompassing a $10.6^{\circ} \times 
10.6^{\circ}$ field of view (FoV) for the detection of UHECRs. 
The telescope is located in front of one of the
fluorescence detectors of the TA experiment.
Since its first operation in 2015, a few campaigns of
joint observations allowed EUSO-TA to detect several UHECRs events as well as
meteors. The limiting magnitude of
5.5 on summed frames ($\sim$3 ms) has been established. Measurements of the UV
night sky emission in different conditions and moon phases have
been performed. 
These observations provided important data to optimise the detector technology 
in view of the upcoming space-based missions. A detailed
description of the results of EUSO-TA and the prospects of future upgrades of 
the detector are described in \cite{piotrowski_icrc19}. Regarding the
performance, 9 UHECR events were identified so far during 120 hours of 
UHECR-dedicated observations which can be directly compared with simulations to
determine the energy threshold of the experiment (see~\cite{bisconti_icrc19}). 
The resulting information is also useful for the upcoming JEM-EUSO balloon and 
space missions. The detector can also be calibrated using stars in the 
FoV, as discussed in~\cite{plebaniak_icrc19}.  
The future upgrades of the detector include a new acquisition system based on
Zynq board 
and the implementation of a self-triggering system already in use for the
balloon missions~\cite{battisti_icrc19}, in order to obtain a detailed 
characterisation of the UHECR detection performances. Moreover, high 
sensitivity CMOS cameras will be used on the same site to detect meteoroids and
search for nuclearites~\cite{kajino_icrc19} in parallel to EUSO-TA 
instrumentation. 

\section{The stratospheric balloon program: EUSO-Balloon, EUSO-SPB1 \& 
EUSO-SPB2}
\label{sec:euso-bal}
The JEM-EUSO program includes three stratospheric balloon missions
with increasing level of performance and upgraded designs. In addition to 
demonstrating the capabilities of the JEM-EUSO instruments to detect and 
reconstruct showers from the edge of space, they also give access to direct 
measurement of the UV nightglow emission and artificial UV contributions above 
ground and oceans, which are important information to optimise the design of 
the space-based missions.
Two balloon flights have been performed so far: EUSO-Balloon (Canada, 1 night) 
and EUSO-SPB1 (Pacific Ocean,
12 nights). A third one (EUSO-SPB2) is under preparation to fly
in 2022 from Wanaka, New Zealand.

EUSO-Balloon~\cite{eusobal} was launched by CNES from
the Timmins base
in Ontario (Canada) on the moonless night of August 25, 2014 UT. After
reaching the floating altitude of $\sim$38 km, EUSO-Balloon imaged the UV
intensity in the wavelength range 290 - 430 nm for more than 5 hours before
descending to ground. The refractor telescope consisted of a similar
apparatus as EUSO-TA. 
The spatial and temporal resolutions of the detector were 130 m and
2.5 $\mu$s,
respectively. The full FoV in nadir mode was $\sim11^{\circ}$.
During 2.5 hours of EUSO-Balloon flight, a
helicopter circled under the balloon operating UV flashers and
a UV laser to simulate the optical signals from UHECRs, to calibrate the
apparatus, and to characterise the optical atmospheric conditions. During
flight EUSO-Balloon took more than 2.5 million images that have been analysed
to infer different information: study of the performance
of the different parts of the detector; response of the detector to the
UV flasher and laser events; UV radiance from the Earth atmosphere and
ground in different conditions: clear and cloudy atmosphere, forests,
lakes, as well as city lights.
The measurement of UV light intensity is relevant for a JEM-EUSO-like mission
as it is one of the key parameters to estimate the exposure curve as a
function of energy \cite{exposure}.
The results of this analysis are reported in~\cite{kenji-uv}
and are in the band of previous measurements confirming a good
understanding of the detector performance also in this respect, which
is very important in view of a space-based observatory like JEM-EUSO.
The helicopter events proved to be extremely useful to understand the system's
performance and to
test the capability of EUSO-Balloon to detect and reconstruct signals similar
to  Extensive-Air-Shower (EAS) events.
Laser tracks were used to test the reconstruction algorithms
\cite{eser-jinst}. 
The data collected by EUSO-Balloon were used together with those collected by
EUSO-TA and experiments as TurLab to define an internal
trigger logic~\cite{trigger} that was implemented for the EUSO-SPB1 flight.

EUSO-SPB1~\cite{spb1}
was launched April 25 2017 from Wanaka New Zealand as a mission of opportunity
on a NASA SPB test flight planned to circle the southern
hemisphere. One of the scientific goals was to make the first
observation of UHECR-EASs by looking down on the atmosphere with an optical
fluorescence detector from the near space altitude of 33 km. 
Unfortunately, although the instrument was showing
nominal behaviour and performances, the flight was terminated prematurely in the
Pacific Ocean about 300 km SE of Easter Island after only 12 days 4 hours 
aloft, due to a leak in the carrying balloon.
The telescope was an upgraded version of that used in the EUSO-Balloon 
mission. In particular, a SiPM camera
was installed~\cite{painter_icrc19}. An autonomous
internal trigger was implemented according to~\cite{trigger} to detect UHECRs.
Prior to flight, in October 2016, the fully assembled EUSO-SPB1 detector was 
tested for a week
at the EUSO-TA site to measure its response and to calibrate it by means of a
portable Ground Laser System (GLS). Observations of Central Laser Facility 
(CLF), stars, meteors were performed.
The $\sim$50\% trigger efficiency was reached at laser energies whose
luminosity is equivalent to $\sim$45$^{\circ}$ inclined EAS of E $\sim3 \times
10^{18}$ eV seen from above by a balloon flying at 33 km altitude.
During flight, $\sim$30 hours of data were collected, the trigger rate was 
tipycally a few Hz, which is compliant 
with JEM-EUSO requirements~\cite{mario-nim2019}. 
A deep analysis of the collected data has been performed. Tracks of
CRs directly crossing the detector have been recognized. However, no
EAS track has been clearly identified~\cite{diaz_icrc19,vrabel_icrc19}. 
Simulations post-flight indicate
that the number of expected events is $\sim$1 in the
available data sample indipendently of the balloon height and UV level
confirming pre-flight expectations for such a flight 
duration~\cite{eser_icrc19}. The role of clouds has been extensively studied
and the exposure reduces to  $\sim$0.6~\cite{kenji_icrc19,silvia_icrc19}
taking them into account.
Data taken by the IR camera of EUSO-SPB1 are being analysed to test methods 
developed in the past~\cite{anzalone_ieee} to retrieve the cloud-top-height
in the frame of the JEM-EUSO mission and to estimate the cloud 
fraction~\cite{anzalone_icrc19}.

The subsequent step of the JEM-EUSO program development is currently under
realization: EUSO-SPB2~\cite{lawrence_icrc19}. It will be equipped with 2 
telescopes. One telescope
will be devoted to UHECR measurements using the fluorescence technique.
The Focal Surface (FS)~\cite{osteria_icrc19} will be equipped with 3 PDMs to 
increase the UHECR collection power.
A more performing optics (Schmidt camera) and 
a shorter temporal resolution of 1 $\mu$s will be used to lower the energy
threshold of the instrument. The FS of the other telescope
will be based on SiPMT sensors and a dedicated electronics to detect
the Cherenkov emission in air by UHECR-generated EASs~\cite{otte_icrc19}. In
perspective they will test the capability to detect EAS generated by
$\nu_{\tau}$ interacting in the Earth crust~\cite{austin_icrc19}. For this 
observation the
detector will be pointing slightly below the limb. The observation above
the limb will allow to study UHECRs through their Cherenkov 
emission~\cite{jacek_icrc19,krolik_icrc19}.
EUSO-SPB2 is programmed to fly by 2022 from Wanaka, New Zealand on a NASA Super
Pressure Balloon.

\section{The precursor space-based missions: TUS and Mini-EUSO}
\label{sec:tus}
The Track Ultraviolet Setup (TUS) detector 
was launched on April 28, 2016 as a
part of the scientific payload of the Lomonosov satellite. TUS~\cite{tus}
is the world's first orbital detector aiming at detecting UHECRs. The
instrument was actively recording data till November 2017. Different
scientific modes were tested: cosmic ray, lightning and meteor modes.
The satellite has a sun-synchronous orbit with an inclination of 97.3$^{\circ}$,
a period of $\sim$94 min, and a height of 470 - 500 km.
The telescope consists of two main parts: a modular Fresnel mirror-concentrator
with an area of $\sim$2 m$^2$ and 256 PMTs
arranged in a 16$\times$16 photo-receiver matrix located in the focal plane of
the mirror. The pixel's FoV is 10 mrad, which corresponds
to a spatial spot of $\sim$5 km $\times$ 5 km at the sea level from a 500 km
orbit height. Thus, the full area observed by TUS at any moment is
$\sim$80 km $\times$ 80 km. TUS is sensitive to the near UV band and in
cosmic ray mode has
a time resolution of 0.8 $\mu$s in a full temporal interval of 256 time steps.
TUS data offer the opportunity to
develop strategies in the analysis and reconstruction of events which will be
essential for future space-based missions such as K-EUSO.
During
its operation TUS has detected about 8$\times$10$^4$ events that have been
subject to an offline analysis to select among them those satisfying
basic temporal and spatial criteria of UHECRs. A few events passed this
first screening. 
One specific event registered in perfect observational conditions was 
deeply scrutinized. Its phenomenology and the possible interpretations are
reported in detail in~\cite{zotov_icrc19}.
In order to perform a deeper analysis of such
candidates, a dedicated algorithm to reconstruct the arrival direction in
TUS events and a specific version of ESAF~\cite{esaf_icrc19} are being developed. 
The data acquired in meteor mode are being analyzed to search for slow events
such as nuclearites. Simulations indicate that nuclearites can be detected in
this mode~\cite{nuclearites_icrc19}, and thanks to the gigantic area that can be
monitored by such space-based detectors, these telescopes have the potential 
to set search yields better than the constraints given by 
former experiments. 

While TUS was conceived mainly to prove the observation of UHECRs from space
with a FS-instrumentation similar to ground-based detectors, 
Mini-EUSO has been developed in order to test the same FS-instrumentation
foreseen for K-EUSO and POEMMA.
Mini-EUSO~\cite{casolino_icrc19} is a UV telescope to be placed in August 2019 
inside the ISS, looking down on the Earth from a nadir-facing window in the 
Russian Zvezda
module. Mini-EUSO will map the Earth in the UV range (290 - 430 nm) with a
spatial and temporal resolutions of $\sim$6 km (similar to TUS) and
2.5 $\mu$s, respectively. Mini-EUSO has a FS similar to EUSO-TA, but with a 
different acquisition system~\cite{minieuso-trigger} .
The optical system consists of 2 Fresnel lenses of 25 cm diameter
with a large FoV of $\sim$22$^{\circ}$.
A multiple level trigger will allow
the measurement of UV transients at different time scales, complementing TUS
observations. Laboratory experiments at TurLab~\cite{miyamoto_icrc19}
as well as open-sky data acquistions with Mini-EUSO engineering model
~\cite{bisconti_miniicrc19} and
simulations confirm the expected sensitivity of Mini-EUSO to 
different light phenomena and to EAS-like 
transients around 10$^{21}$ eV. It is planned to use a ground-based UV laser system to test such
sensitivity~\cite{kungel_icrc19}. Mini-EUSO has also other scientific 
objectives, among them the study of Transient Luminous Events (TLEs), 
meteors, and the search for 
stranglets and detection of space debris~\cite{ebisu,hirokodeb_icrc19}.

\section{The target space-based missions of the program: K-EUSO \& POEMMA}
\label{sec:keuso}
The central objective of K-EUSO~\cite{keuso} is the first consistent 
measurement of the UHECR flux over the entire sky with unprecedented 
and almost uniform exposure.
K-EUSO is a result of the joint efforts to improve the performance of the
Russian KLYPVE mission~\cite{klypve}, by employing the technologies developed for the
JEM-EUSO mission, such as the focal surface detectors and the readout
electronics.
Since its first conception as KLYPVE, K-EUSO project has passed various
modifications aimed to increase FoV and UHECR statistics.
It will be the first detector with a real capability for UHECR spectrum and
anisotropy study with a yearly exposure of 
$\sim$4 times Auger and a flat full celestial sphere
coverage. The adopted optical layout is a Schmidt camera covering a
FoV of 40 degree with an entrance pupil diameter of 2.5 m, a 4 m diameter
spherical mirror and a focal length of 1.7 m.
The temporal (sampling time is 1$\mu$s) and spatial (angular resolution per 
pixel 0.066$^{\circ}$) evolution of UV light recorded by K-EUSO will allow
the reconstruction of the EAS, with sufficient energy and arrival direction
resolutions.
The camera focal plane consists of 52 PDMs like in EUSO-SPB2, for a total of
1.2$\times$10$^5$ pixels. A pixel will cover $\sim$0.8 km on the surface of the
 Earth for ISS altitude of 400 km.
Attached to the Russian MRM-1 module on-board ISS, K-EUSO is planned to operate
for minimum of 2 years and it can function more than 6 years if the lifetime of
 the ISS is extended.

The Probe Of Extreme Multi-Messenger Astrophysics (POEMMA)
mission~\cite{poemma} is being designed to establish charged particle astronomy
with UHECRs and to observe cosmic neutrinos from space. 
POEMMA will monitor colossal volumes of the Earth's atmosphere to
detect extensive air showers (EASs) produced by extremely energetic cosmic 
messengers: cosmic neutrinos above 20 PeV and UHECRs above 20 EeV over the 
entire sky.
The POEMMA design combines the concept developed for the OWL mission
~\cite{owl} and the experience of the JEM-EUSO fluorescence
detection camera. POEMMA is composed of two identical satellites flying in
formation at 525 km altitude with the ability to observe overlapping regions
during moonless nights at angles ranging from Nadir to just above the limb of
the Earth, but also with independent pointing strategies to exploit at maximum
the scientific program of the mission~\cite{angela_icrc19}.
Each POEMMA satellite consists of a 4-meter photometer
designed with Schmidt wide (45$^\circ$) FoV optics.
The POEMMA FS is composed of a hybrid of two types of cameras: about 90\%
of the FS is dedicated to the POEMMA fluorescence camera (PFC), while POEMMA
Cherenkov camera (PCC) occupies the crescent moon shaped edge of the
FS which images the limb of the Earth. The PFC is composed of JEM-EUSO
PDMs based on MAPMTs. The typical time between images for the PFC is about
1 $\mu$sec. The much faster POEMMA Cherenkov camera (PCC) is composed of
Silicon photo-multipliers (SiPMs) also flown
in EUSO-SPB1 and soon to be tested in space with Mini-EUSO. The PFC registers
UHECR tracks from Nadir to just below the Earth's limb, while the PCC
registers light within the Cherenkov emission cone of up-going showers around
the limb of the Earth and also from high energy cosmic rays above the limb of
the Earth.

\section{Acknowledgments}
This work was partially supported by Basic Science Interdisciplinary Research
Projects of RIKEN and JSPS KAKENHI Grant (JP17H02905, JP16H02426 and 
JP16H16737), by the Italian Ministry of Foreign Affairs and International 
Cooperation, by the Italian Space Agency through the ASI INFN agreement 
n. 2017-8-H.0 and contract n. 2016-1-U.0, by NASA award 11-APRA-0058 in the USA, by the
Deutsches Zentrum f\"ur Luft- und Raumfahrt, by the French space agency
CNES, the Helmholtz Alliance for Astroparticle Physics funded by the
Initiative and Networking Fund of the Helmholtz Association (Germany), by
Slovak Academy of Sciences MVTS JEMEUSO as well as VEGA grant agency project
2/0132/17,
by National Science Centre in Poland grant (2015/19/N/ST9/03708 and 
2017/27/B/ST9/02162), by Mexican funding agencies PAPIIT-UNAM, CONACyT and the 
Mexican Space Agency (AEM). Russia is supported by ROSCOSMOS and the Russian 
Foundation for
Basic Research Grant No 16-29-13065. Sweden is funded by the Olle Engkvist 
Byggm\"astare Foundation.


\begin{thebibliography}{99}


\bibitem{EUSO-Program}
  M.~Ricci (JEM-EUSO Coll.),
  J.\ Phys.\ Conf.\ Ser.\  {\bf 718} (2016) no.5,  052034.

\bibitem{jemeuso}
 J.H. Adams {\it et al} (JEM-EUSO Coll.), Experimental Astronomy
{\bf 40} (2015) 3.

\bibitem{eusota}
 G. Abdellaoui {\it et al} (JEM-EUSO Coll.), Astroparticle Physics
{\bf 102} (2018) 98.

\bibitem{eusobal}
 J.H. Adams {\it et al} (JEM-EUSO Coll.), Experimental Astronomy
{\bf 40} (2015) 281.

\bibitem{spb1}
 L. Wiencke and A. Olinto for the JEM-EUSO Coll., PoS(ICRC2017), {\bf 1097}.

\bibitem{spb2}
 J.H. Adams {\it et al}, ArXiv e-prints [[arXiv]1703.04513 (2017).

\bibitem{tus}
 P. Klimov {\it et al} (TUS Coll.), Space Science Reviews
{\bf 8} (2017) 1.

\bibitem{minieuso}
 F. Capel {\it et al}, Advances in Space Research, {\bf 62} (2018) 2954.

\bibitem{keuso}
 M. Casolino M. {\it et al} (JEM-EUSO Coll.), PoS(ICRC2017) {\bf 368}.

\bibitem{poemma}
 A. Olinto {\it et al} (POEMMA Coll.), PoS(ICRC2017) {\bf 542}.

\bibitem{fenu_icrc19}
F. Fenu {\it et al} (JEM-EUSO Coll.), {\it Results from the first missions
of the JEM-EUSO program}, PoS(ICRC2019) {\bf 251}.

\bibitem{piotrowski_icrc19}
L. Piotrowski {\it et al} (JEM-EUSO Coll.), {\it Results and status of the 
EUSO-TA detector},  PoS(ICRC2019) {\bf 388}.

\bibitem{bisconti_icrc19}
F. Bisconti {\it et al} (JEM-EUSO Coll.), {\it EUSO-TA ground based 
fluorescence detector: analysis of the detected events}, PoS(ICRC2019) {\bf 197}.

\bibitem{plebaniak_icrc19}
Z. Plebaniak {\it et al} (JEM-EUSO Coll.), {\it Calibration of EUSO-TA detector with stars}, PoS(ICRC2019) {\bf 393}.

\bibitem{battisti_icrc19}
M. Battisti {\it et al} (JEM-EUSO Coll.), {\it EUSO-TA \& EUSO-SPB2 trigger 
logic}, PoS(ICRC2019) {\bf 426}.

\bibitem{kajino_icrc19}
R. Ide {\it et al}, {\it Study of Fast Moving Nuclearites and Meteoroids using 
High Sensitivity CMOS Camera}, PoS(ICRC2019) {\bf 525}.

\bibitem{exposure}
J.H. Adams {\it et al} (JEM-EUSO Coll.),  Astroparticle Physics {\bf 44}
(2013) 76.

\bibitem{kenji-uv}
 G. Abdellaoui {\it et al} (JEM-EUSO Coll.), Astroparticle Physics
{\bf 111} (2019) 54.

\bibitem{eser-jinst}
G. Abdellaoui {\it et al} (JEM-EUSO Coll.), J. of Instrumentation
{\bf 13} (2018) 05023.

\bibitem{trigger}
G. Abdellaoui {\it et al} (JEM-EUSO Coll.), Nucl. Instr. \& Meth. A
{\bf 866} (2017) 150.

\bibitem{painter_icrc19}
W. Painter {\it et al} (JEM-EUSO Coll.), {\it Silicon Photomultipliersfor 
Orbital Ultra High Energy Cosmic Ray Observation}, PoS(ICRC2019) {\bf 285}.

\bibitem{mario-nim2019}
 M. Battisti {\it et al} (JEM-EUSO Coll.), Nucl. Instr. \& Meth. A
{\bf 936} 349 (2019).

\bibitem{diaz_icrc19}
A. Diaz {\it et al} (JEM-EUSO Coll.), {\it EUSO-SPB1: Flight data classification
 and Air shower search results}, PoS(ICRC2019) {\bf 240}.

\bibitem{vrabel_icrc19}
M. Vrabel {\it et al} (JEM-EUSO Coll.), {\it Machine Learning Approach for Air 
Shower Recognition in the EUSO-SPB Experiment Data}, PoS(ICRC2019) {\bf 456}.

\bibitem{eser_icrc19}
J. Eser {\it et al} (JEM-EUSO Coll.), {\it EUSO-SPB1 overview}, PoS(ICRC2019)
{\bf 247}.

\bibitem{kenji_icrc19}
K. Shinozaki {\it et al} (JEM-EUSO Coll.), {\it An estimation of the exposure of air shower detectionby the EUSO-SPB1 mission}, PoS(ICRC2019) {\bf 427}.

\bibitem{silvia_icrc19}
S. Monte {\it et al} (JEM-EUSO Coll.), {\it WRF and radiative methods for Cloud
Top Height retrieval along EUSO-SPB1 trajectory}, PoS(ICRC2019) {\bf 455}.

\bibitem{anzalone_ieee}
 A. Anzalone {\it et al}, IEEE Trans. Geoscience \& Remote Sensing
{\bf 57} 304 (2019).

\bibitem{anzalone_icrc19}
A. Bruno {\it et al} (JEM-EUSO Coll.), {\it A method for Cloud Mapping in the
 Field of View of the Infra-Red Camera during the EUSO-SPB1 flight}, 
PoS(ICRC2019) {\bf 454}.

\bibitem{lawrence_icrc19}
L. Wiencke {\it et al} (JEM-EUSO Coll.), {\it The Extreme Universe Space 
Observatory on a Super-Pressure Balloon II Mission}, PoS(ICRC2019) {\bf 466}.

\bibitem{osteria_icrc19}
V. Scotti {\it et al} (JEM-EUSO Coll.), {\it The Data Processor of the 
EUSO-SPB2 telescopes}, PoS(ICRC2019) {\bf 368}.

\bibitem{otte_icrc19}
N. Otte {\it et al} (JEM-EUSO Coll.), {\it Development of a Cherenkov Telescope
 for the Detection of Ultra-High Energy Neutrinos with EUSO-SPB2 and POEMMA}, 
PoS(ICRC2019) {\bf 977}.

\bibitem{austin_icrc19}
A. Cummings {\it et al}, {\it A More Complete Phenomenology of
 Tau Lepton Induced Air Showers}, PoS(ICRC2019) {\bf 862}.

\bibitem{jacek_icrc19}
J. Szabelski {\it et al} (JEM-EUSO Coll.), {\it Cosmic Ray Mass Composition 
around 10$^{18}$ eV with Horizontal Cherenkov EAS Light Balloon
 Measurements}, PoS(ICRC2019) {\bf 433}.

\bibitem{krolik_icrc19}
K. Krolik {\it et al} (JEM-EUSO Coll.), {\it Cherenkov light from horizontal Air
 Showers}, PoS(ICRC2019) {\bf 321}.

\bibitem{zotov_icrc19}
M. Zotov {\it et al} (TUS Coll.), {\it An UHECR-like event registered with the 
TUS orbital detector}, PoS(ICRC2019) {\bf 193}.

\bibitem{esaf_icrc19}
F. Fenu {\it et al} (JEM-EUSO Coll.), {\it Simulations for the JEM-EUSO program
with ESAF} PoS(ICRC2019) {\bf 252}.

\bibitem{nuclearites_icrc19}
K. Shinozaki {\it et al} (TUS Coll.), {\it Search for slow-moving 
nuclearite-induced events using the TUS orbital UHECR detector} PoS(ICRC2019)
{\bf 545}.

\bibitem{casolino_icrc19}
M. Casolino {\it et al} (JEM-EUSO Coll.), {\it The MINI-EUSO mission to study 
UV emissions from the International Space Station}, PoS(ICRC2019) {\bf 212}.


\bibitem{minieuso-trigger}
 A. Belov {\it et al}, Advances in Space Research, {\bf 62} (2018) 2966.

\bibitem{miyamoto_icrc19}
H. Miyamoto {\it et al} (JEM-EUSO Coll.), {\it The EUSO@Turlab Project: Tests 
of Mini-EUSO Engineering Model}, PoS(ICRC2019) {\bf 194}.

\bibitem{bisconti_miniicrc19}
F. Bisconti {\it et al} (JEM-EUSO Coll.), {\it Mini-EUSO engineering model: 
tests in open-sky condition}, PoS(ICRC2019) {\bf 198}.

\bibitem{kungel_icrc19}
V. Kungel {\it et al} (JEM-EUSO Coll.), {\it UV laser system test of Mini-EUSO},
PoS(ICRC2019) {\bf 325}.

\bibitem{ebisu}
 T. Ebisuzaki et al., Acta Astronautica {\bf 112} (2015) 102.

\bibitem{hirokodeb_icrc19}
H. Miyamoto {\it et al} (JEM-EUSO Coll.), {\it Space debris detection and 
tracking with the techniques of cosmic ray physics}, PoS(ICRC2019) {\bf 253}.

\bibitem{klypve}
B. Khrenov et al., Nuclear Physics B Proceedings
Supplements, {\bf 113/1} (2002) 115.

\bibitem{owl}
F.W. Stecker {\it et al}, Nucl. Phys. B {\bf 136C} (2004) 433.

\bibitem{angela_icrc19}
A. Olinto {\it et al} (POEMMA Coll.), {\it POEMMA: Probe Of Extreme 
Multi-Messenger Astrophysics}, PoS(ICRC2019) {\bf 378}.

\end{thebibliography}
\end{document}